\documentclass[sigconf]{acmart}
\AtBeginDocument{
  }

\usepackage{enumitem}
\usepackage{tabularx}  
\usepackage{booktabs}
\usepackage{array}     
\usepackage[most]{tcolorbox} 
\tcbuselibrary{skins}        
\tcbuselibrary{breakable}   
\usepackage{balance}

\usepackage[svgnames]{xcolor}
\definecolor{appendixqbg}{HTML}{E3F2FD}    
\definecolor{appendixqframe}{HTML}{90CAF9} 
\definecolor{appendixqtitlebg}{HTML}{90CAF9}
\definecolor{appendixsubbg}{HTML}{FCFCFC}      
\definecolor{appendixsubframe}{HTML}{E0E0E0}   

\newtcolorbox{questionboxappendix}[1]{
    enhanced, 
    breakable,
    colback=appendixqbg,
    colframe=appendixqframe,
    boxrule=0.8pt, 
    arc=2mm,
    fonttitle=\bfseries,
    coltitle=black, 
    title=#1,
    attach boxed title to top left={yshift=-2mm-\tcboxedtitleheight/2,xshift=3mm},
    boxed title style={
        colframe=appendixqframe, 
        colback=appendixqtitlebg,
        arc=2mm, 
        outer arc=2mm,
        bottomrule=0pt, toprule=0pt, leftrule=0pt, rightrule=0pt,
    },
    top=5mm, 
    before=\par\medskip\noindent,
    after=\par\medskip
}

\newtcolorbox{subblockappendix}[1]{
    breakable,
    colback=appendixsubbg,
    colframe=appendixsubframe,
    boxrule=0.5pt, arc=1mm,
    title=#1,
    fonttitle=\bfseries\normalsize,
    coltitle=DarkSlateGray,
    top=2mm, bottom=2mm, left=2mm, right=2mm,
    before=\par\smallskip\noindent,
    after=\par\smallskip
}

\copyrightyear{2026}
\acmYear{2026}
\setcopyright{cc}
\setcctype{by-nc-nd}
\acmConference[KDD 2026] {Proceedings of the 32nd ACM SIGKDD Conference on Knowledge Discovery and Data Mining V.2}{August 9--13, 2026}{Jeju Island, Republic of Korea.}
\acmBooktitle{Proceedings of the 32nd ACM SIGKDD Conference on Knowledge Discovery and Data Mining V.2 (KDD 2026), August 9--13, 2026, Jeju Island, Republic of Korea}
\acmISBN{979-8-4007-2259-2/2026/08}
\acmDOI{10.1145/3770855.3817512}

\begin{document}

\title{DMind Benchmark: Toward a Holistic Assessment of LLM Capabilities across the Web3 Domain}

\copyrightyear{2026}
\acmYear{2026}
\setcopyright{cc}
\setcctype{by-nc-nd}
\acmConference[KDD '26]{Proceedings of the 32nd ACM SIGKDD Conference on Knowledge Discovery and Data Mining V.2}{August 09--13, 2026}{Jeju Island, Republic of Korea}
\acmBooktitle{Proceedings of the 32nd ACM SIGKDD Conference on Knowledge Discovery and Data Mining V.2 (KDD '26), August 09--13, 2026, Jeju Island, Republic of Korea}
\acmDOI{10.1145/3770855.3817512}
\acmISBN{979-8-4007-2259-2/2026/08}

\author{Enhao Huang}
\authornote{Also with The State Key Laboratory of Blockchain and Data Security.}
\affiliation{%
  \institution{Zhejiang University}
  \city{Hangzhou}
  \country{China}}
\email{huangenhao@zju.edu.cn}

\author{Pengyu Sun}
\authornotemark[1]
\affiliation{%
  \institution{Zhejiang University}
  \city{Hangzhou}
  \country{China}}
\email{3220105108@zju.edu.cn}

\author{Shuxun Wang}
\authornotemark[1]
\affiliation{%
  \institution{Zhejiang University}
  \city{Hangzhou}
  \country{China}}
\email{shuxunwang@zju.edu.cn}

\author{Zixin Lin}
\affiliation{%
  \institution{Zhejiang University}
  \city{Hangzhou}
  \country{China}}
\email{linzixin@zju.edu.cn}

\author{Alex Chen}
\affiliation{%
  \institution{Zhejiang University}
  \city{Hangzhou}
  \country{China}}
\email{cq1757@zju.edu.cn}

\author{Kaichun Hu}
\authornotemark[1]
\affiliation{%
  \institution{Zhejiang University}
  \city{Hangzhou}
  \country{China}}
\email{tauhkc@zju.edu.cn}

\author{Joey Ouyang}
\affiliation{%
  \institution{DMind.ai}
  \country{Singapore}}
\email{knight@dmind.ai}

\author{Frank Li}
\affiliation{%
  \institution{DMind.ai}
  \country{Singapore}}
\email{frank@dmind.ai}

\author{Zhiyu Zhang}
\affiliation{%
  \institution{Zhejiang University}
  \city{Hangzhou}
  \country{China}}
\email{zy-zhang@zju.edu.cn}

\author{Haobo Wang}
\affiliation{%
  \institution{Zhejiang University}
  \city{Hangzhou}
  \country{China}}
\email{wanghaobo@zju.edu.cn}

\author{Yiming Li}
\authornote{Corresponding authors.}
\affiliation{%
  \institution{Nanyang Technological University}
  \country{Singapore}}
\email{liyiming.tech@gmail.com}

\author{Zhan Qin}
\authornotemark[1]
\affiliation{%
  \institution{Zhejiang University}
  \city{Hangzhou}
  \country{China}}
\email{qinzhan@zju.edu.cn}

\author{James Yi}
\affiliation{%
  \institution{DMind.ai}
  \country{Singapore}}
\email{james@dmind.ai}

\author{Gang Zhao}
\affiliation{%
  \institution{DMind.ai}
  \country{Singapore}}
\email{garry@dmind.ai}

\author{Ziang Ling}
\affiliation{%
  \institution{DMind.ai}
  \country{Singapore}}
\email{tony@dmind.ai}

\author{Lowes Yang}
\authornote{Corresponding authors.}
\affiliation{%
  \institution{DMind.ai}
  \country{Singapore}}
\email{lowesyang@dmind.ai}

\renewcommand{\shortauthors}{Enhao Huang et al.}

\begin{abstract}
The Web3 ecosystem, underpinned by cryptographic primitives and decentralized consensus, represents a high-stakes environment where software vulnerabilities and incentive misalignments translate directly into financial loss. As Large Language Models (LLMs) are increasingly integrated into this domain for tasks ranging from smart contract auditing to decentralized finance analytics, ensuring their reliability is paramount. However, general-purpose benchmarks fail to capture the specialized reasoning required for these adversarial and protocol-driven settings. To bridge this gap, we introduce \textbf{DMind Benchmark}, a comprehensive evaluation suite designed to rigorously assess LLM proficiency across the Web3 stack. DMind Benchmark encompasses nine distinct subdomains (spanning infrastructure, smart contracts, token economics, etc.) and combines objective knowledge retrieval with complex open-ended reasoning tasks that emulate real-world operational challenges. We conduct an extensive evaluation of 31 leading proprietary and open-weights models, employing a contamination-aware pipeline and verifying the statistical robustness of our scoring protocol through rigorous cross-judge consistency checks. Our analysis reveals a critical dichotomy: while models demonstrate competence in foundational infrastructure concepts, they exhibit significant vulnerabilities in high-reasoning tasks such as security auditing. Furthermore, we provide a Pareto analysis to guide cost-effective deployment and demonstrate through adversarial experiments that high performance on DMind Benchmark necessitates genuine reasoning rather than superficial memorization. Since its open-source release in April 2025, DMind Benchmark achieved the \#1 trending position on Hugging Face for nearly a week and accumulated over 13k downloads by June 2026, establishing itself as a standard for advancing secure and trustworthy AI in Web3. The dataset and evaluation toolkit are publicly available at \url{https://huggingface.co/datasets/DMindAI/DMind_Benchmark}.
\end{abstract}

\ccsdesc[500]{Computing methodologies~Language resources}

\keywords{Large Language Models, Benchmarking, Evaluation Methodologies, Web3, Security and Privacy, Blockchain}
\begin{teaserfigure}
\centering
  \includegraphics[width=0.8\textwidth]{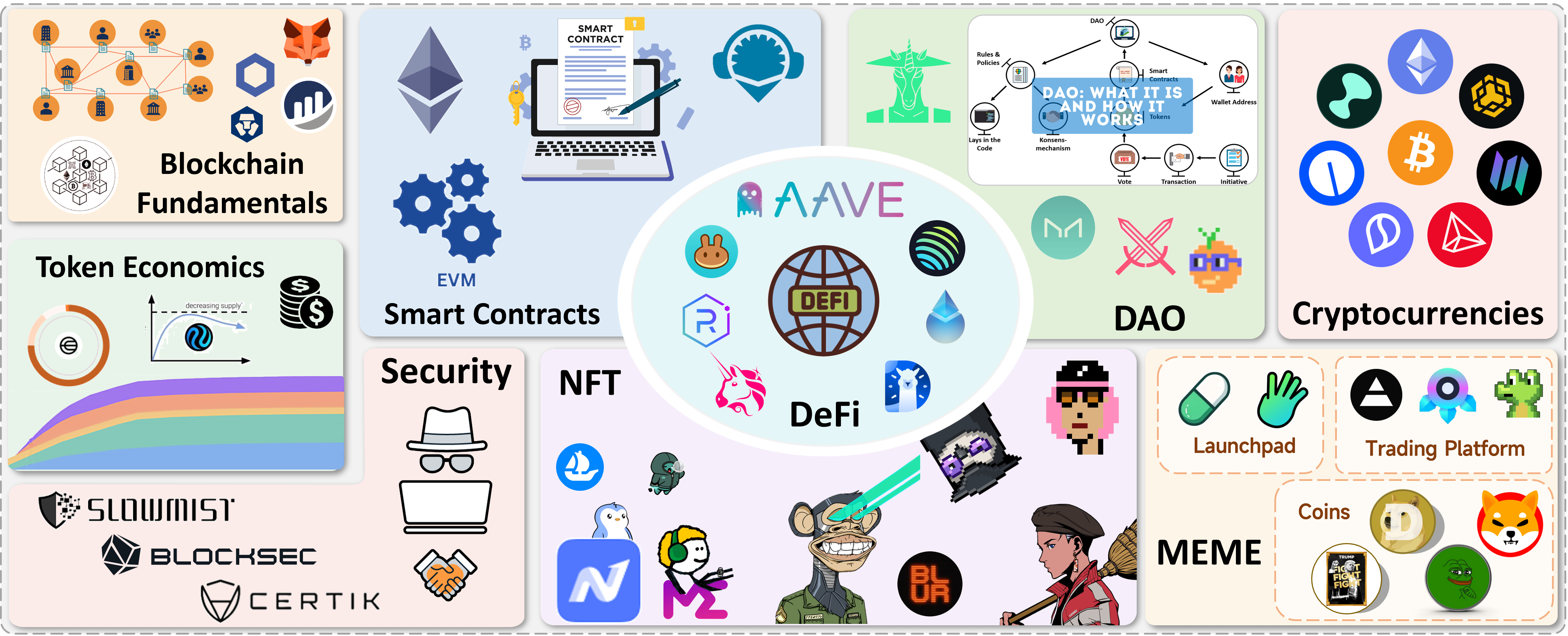}
\caption{Overview of the nine key Web3 subdomains covered and evaluated by the \textbf{DMind Benchmark}}
\Description{Overview of the nine key Web3 subdomains evaluated in the DMind Benchmark, showing their interconnections within the broader Web3 ecosystem.}
  \label{fig:teaser}
\end{teaserfigure}

\maketitle

\section{Introduction}

Large Language Models (LLMs) have evolved from fluent text generators into general-purpose systems that can follow complex instructions and perform multi-step reasoning \cite{openai2023gpt4technicalreport, deepseekai2025deepseekr1incentivizingreasoningcapability, touvron2023llamaopenefficientfoundation, gemmateam2024gemmaopenmodelsbased}. As these models become production-ready, they are increasingly embedded in domain-critical workflows such as biomedical decision support \cite{singhal2023large}, financial analysis \cite{wu2023bloomberggpt}, and legal reasoning \cite{essay104811}. In these settings, broad language ability is necessary but not sufficient. Practical reliability requires precise domain knowledge, faithful numerical and logical reasoning, and robustness under ambiguity and adversarial inputs. As a result, domain-grounded evaluation has become a bottleneck for responsible deployment.

In this paper, we study LLMs in the Web3 domain. Web3 seeks to re-architect the internet around decentralized ownership and trust minimization, replacing centralized intermediaries with cryptographic primitives and distributed consensus \cite{buterin2013ethereum, wood2014ethereum, chatter2025efficientofvarious, geren2025blockchainforlarge}. Modern Web3 systems span smart contracts, DeFi protocols \cite{Ozili2022}, NFTs \cite{wang2021non}, DAOs \cite{Bellavitis03042023}, on-chain governance, and privacy-enhancing infrastructures, all underpinned by cryptoeconomic incentives and an adversarial security environment. Web3 therefore combines software engineering with mechanism design, and it operates under unusually high stakes. Applications often manage real capital, are difficult to patch once deployed, and face constant attack. Even small reasoning errors can translate into exploitable contracts, misaligned incentives, or governance failures. These properties make Web3 a demanding setting for assessing whether LLMs can be trusted as assistants for developers, auditors, and end users.

LLMs are already being integrated into Web3 workflows. Recent studies explore LLM-assisted smart contract development and refactoring \cite{nijkamp2023codegen, nam2024using, zhong2024can}, vulnerability understanding and audit support \cite{wu2024semantic}, and tooling for documentation and user support \cite{suri2024docedit, dearstyne2024supporting}. Beyond code, LLMs are used for blockchain data analytics \cite{toyoda2024blockchain}, DeFi interaction agents \cite{mothukuri2024ai}, and market-oriented tasks such as cryptocurrency forecasting \cite{li2024reflective}. Despite this momentum, progress is constrained by the lack of a shared, domain-grounded evaluation standard. General-purpose benchmarks such as MMLU \cite{hendrycks2021measuring}, BIG-Bench \cite{srivastava2025beyond}, and HELM \cite{liang2023holistic} provide valuable coverage of broad capabilities, but they do not directly test Web3-specific competencies such as EVM-level debugging, multi-step exploit reasoning, or incentive analysis under strategic behavior. In contrast to medicine and finance, where domain benchmarks have become a foundation for systematic progress \cite{chen2024finben, jin2024disease}, the Web3 ecosystem has lacked a comprehensive and practical evaluation suite.

To address this gap, we introduce the \textbf{DMind Benchmark} \cite{huang2025dmind}, a holistic evaluation suite designed to assess LLM proficiency across the Web3 stack. DMind Benchmark is organized into nine domains: \emph{(1) blockchain fundamentals, (2) blockchain infrastructure, (3) smart contracts, (4) DeFi mechanisms, (5) DAOs, (6) NFTs, (7) token economics, (8) meme concepts, and (9) security vulnerabilities}. We adopt this taxonomy for a simple reason: it mirrors how Web3 systems are built and how they succeed or fail in the real world. One lens follows the engineering path of a typical system, from shared concepts (fundamentals), to the execution environment and scalability constraints (infrastructure), to the programmable layer where behavior is encoded (smart contracts), and then to the dominant application families seen in practice (DeFi, DAOs, NFTs). A second lens captures the forces that shape outcomes even when code appears correct. Token economics formalizes incentives and strategic behavior; meme concepts capture fast-moving narratives that affect adoption, attention, and coordination; security vulnerabilities represent the adversarial pressure that stress tests every layer. Together, these nine domains offer compact coverage with clear diagnostic value, enabling evaluation that is both end-to-end and actionable.

The construction of DMind Benchmark employs a rigorous pipeline characterized by strict provenance tracking and contamination awareness. We curate a whitelist of authoritative Web3 sources, preserve metadata for auditability, and apply sanitization, deduplication, and privacy redaction before expert review. Domain specialists convert high-value concepts into evaluation items using paraphrasing and option randomization to reduce shortcut learning. DMind Benchmark combines objective items that test precise knowledge with open-ended tasks that reflect operational demands, including smart contract debugging, numerical reasoning over on-chain data, and security auditing. For these subjective tasks, we use a decomposed rubric that breaks each problem into weighted atomic criteria, strongly penalizing hallucinations while rewarding precise and actionable reasoning. We validate the resulting scores through cross-judge consistency experiments, showing that conclusions are stable across evaluator choices.

Using DMind Benchmark, we evaluate \textbf{31} prominent LLMs spanning proprietary and open-weights families. Our study provides an overall performance ranking with multi-run stability, a judge-robust analysis of subjective scoring, and a subdomain diagnosis that localizes capability gaps. We find that current models are relatively strong on infrastructure-related knowledge, but they exhibit persistent weaknesses in token economics and security. We further analyze cost versus performance to identify Pareto-efficient deployment choices, and we run an adversarial fine-tuning experiment showing that naive memorization yields negligible gains, suggesting that high scores require genuine reasoning.

Our primary contributions are:
\begin{itemize}[leftmargin=*]
    \item We introduce \textbf{DMind}, a comprehensive Web3 benchmark evaluating foundational knowledge and realistic, high-stakes reasoning tasks across nine core domains. Since its open-source release in April 2025, DMind sustained the \#1 position on the Hugging Face trending list for nearly a week and exceeded 13k downloads by June 4, 2026.
    \item We propose a hybrid evaluation protocol that cleanly separates objective assessment from rubric-based scoring of subjective Web3 tasks, and we demonstrate strong robustness through extensive ablations and cross-judge consistency checks.
    \item We provide a large-scale evaluation of 31 leading LLMs with a fine-grained capability diagnosis, an efficiency analysis of cost versus performance, and evidence of contamination resistance, offering practical guidance for developing and deploying reliable LLM systems for Web3 applications.
\end{itemize}

\begin{figure*}[t!]
    \centering
    \includegraphics[width=0.86\textwidth]{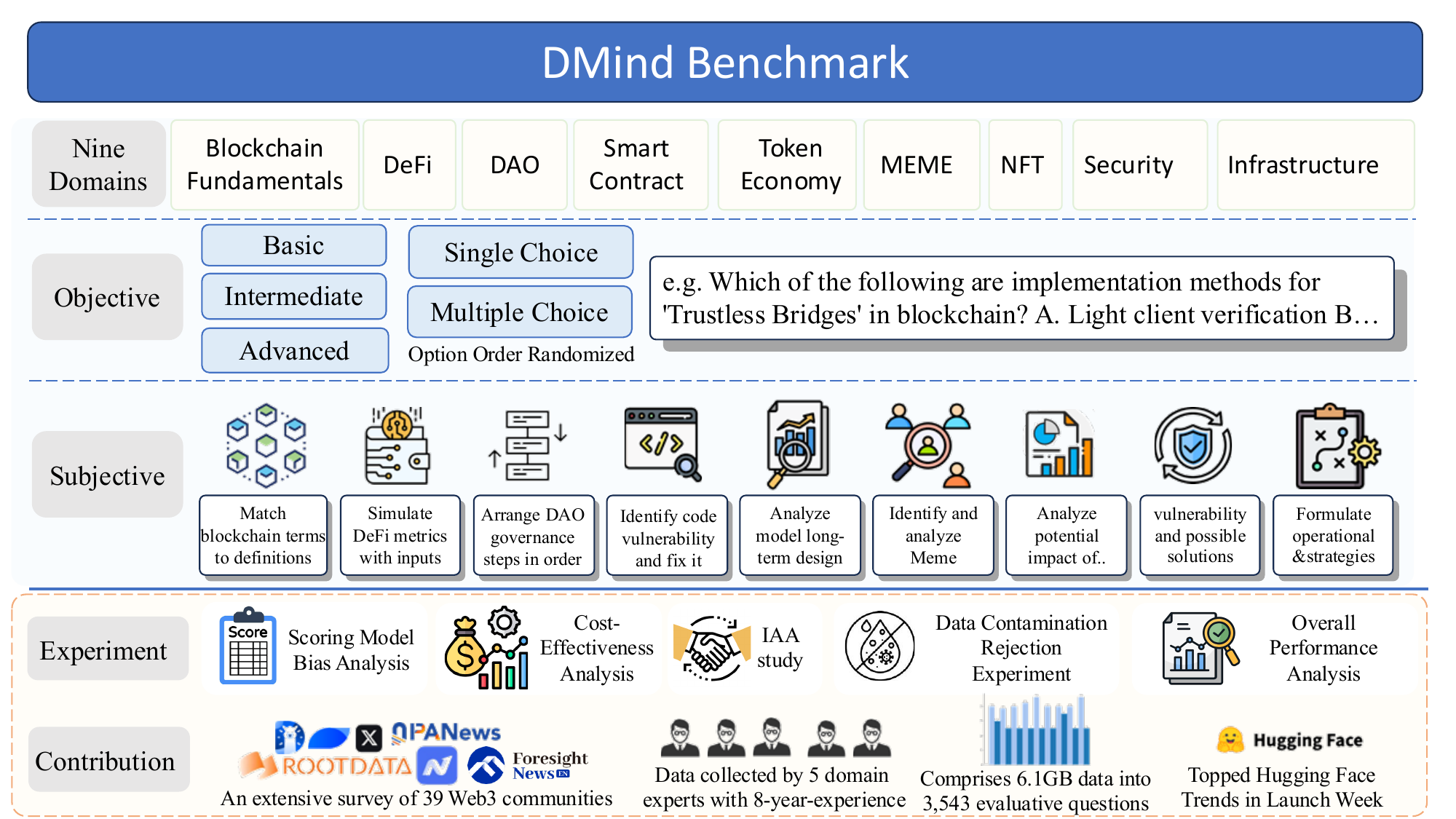}
    \caption{The DMind Benchmark framework, illustrating its nine evaluated Web3 domains, diverse objective and subjective task structures, and key metrics related to its development and community impact.}
    \label{fig:pipeline}
    \vspace{-0.8em}
\end{figure*}

\section{Related Work}
\label{sec:related_work}

\subsection{LLM Evaluation Benchmarks}
\label{sec:llm_evaluation_benchmarks}
Evaluating the capabilities of Large Language Models (LLMs) has garnered significant attention, leading to numerous benchmarks assessing different facets of model performance. Early general-purpose benchmarks like GLUE \cite{wang2018glue} and SuperGLUE \cite{wang2019superglue} focused primarily on natural language understanding. More recent and comprehensive efforts, including MMLU \cite{hendrycks2021measuring}, BIG-Bench \cite{srivastava2025beyond}, and HELM \cite{liang2023holistic}, provide broader assessments of advanced capabilities such as higher-level reasoning, domain knowledge, and instruction-following proficiency. MMLU evaluates models across 57 diverse subject areas; BIG-Bench incorporates over 200 tasks designed to probe aptitudes beyond conventional NLP benchmarks; and HELM offers a framework to assess multiple dimensions like accuracy, calibration, robustness, fairness, and efficiency.

While these general benchmarks offer invaluable insights, they often do not explicitly address the specialized demands of niche domains. This limitation has spurred the creation of domain-specific benchmarks to rigorously evaluate models in specialized areas. For instance, in the medical field, MedQA \cite{jin2024disease}, MultiMedQA \cite{singhal2023large}, and MedMCQA \cite{pal2022medmcqa} examine medical knowledge and diagnostic reasoning. Similarly, finance has seen benchmarks like FinBen \cite{chen2024finben} and FinEval \cite{zhang2025fineval} for assessing the understanding of financial concepts and analytical capabilities. Other notable examples include LegalBench \cite{guha2023legalbench} for legal reasoning, CyberBench \cite{wang2024cyberbench} for cybersecurity knowledge, and SafetyBench \cite{zhang2024safetybench} for evaluating model safety in critical scenarios. Recently, further efforts have emerged to stress-test model robustness and safety, such as RAS-Eval \cite{fu2025ras} for real-world agent security evaluation and TRIDENT \cite{hui2025trident} for benchmarking safety in finance, medicine, and law. Despite these advancements, to the best of our knowledge, a benchmark specifically for evaluating LLM capabilities within the Web3 domain—characterized by its technical intricacies, interdisciplinary nature, and critical security considerations—has been notably absent.

\subsection{Web3 Technologies and Applications} \label{sec:web3_technologies_condensed}
Web3 represents a shift to a decentralized ecosystem built on blockchain, emphasizing user control and trustlessness \cite{yli2016where,yaga2018blockchain}. This section traces Web3's development, highlighting key milestones in its infrastructure, applications, governance, and security.

Web3's foundations began with Nick Szabo's 1997 concept of smart contracts for automating agreements \cite{szabo1997formalizing}. In 2008, Satoshi Nakamoto's Bitcoin introduced distributed ledgers and cryptographic consensus, creating the trustless backbone for Web3 infrastructures \cite{nakamoto2008bitcoin}. In 2014, Vitalik Buterin's Ethereum enabled programmable dApps \cite{buterin2014ethereum}, and Gavin Wood coined the term Web3 for this decentralized ecosystem \cite{wood2014web3}. By 2017, empirical analyses revealed smart contract vulnerabilities \cite{bartoletti2017empirical}, prompting the development of tools for security and efficiency \cite{liu2018reguard,lai2020static,saha2021dhacs}.
The 2020s addressed scalability with Layer-1 and Layer-2 solutions, improving interoperability \cite{belchior2021survey,zhou2020solutions}. This spurred the growth of dApps like Decentralized Finance (DeFi) for trustless financial services \cite{chen2020blockchain,werner2022sok} and Non-Fungible Tokens (NFTs) for unique digital assets \cite{wang2021non,nadini2021mapping}.
Governance evolved with Decentralized Autonomous Organizations (DAOs) around 2019, enabling community-led initiatives through token-voting \cite{wang2019decentralized,hassan2021decentralized}. Tokenomics began shaping incentives \cite{ito2024cryptoeconomics,catalini2022some}, while meme-driven trends spurred adoption \cite{long2025coinclip,krause2024beyond}.
Security measures evolved to counter threats like flash loan exploits and Sybil attacks \cite{islam2021review}, with audits and formal verification preserving system integrity.

These milestones highlight Web3’s interdisciplinary nature, combining cryptography, distributed systems, and economics. Effective modeling in this domain requires language understanding to synthesize its technical, financial, and social concepts.

\subsection{LLMs for Web3 Applications}
Recent studies highlight the strides LLMs are making in empowering the Web3 domain \cite{luo2024bc4llm}. Notably, they are enhancing smart contract security through improved vulnerability detection \cite{wu2024semantic} and accelerating development via automated code generation \cite{nijkamp2023codegen, nam2024using, zhong2024can}, while also streamlining documentation support \cite{suri2024docedit, dearstyne2024supporting}. Furthermore, LLMs are offering deeper insights via blockchain data analytics \cite{toyoda2024blockchain}, aiding in cryptocurrency price forecasting \cite{li2024reflective}, and enabling more intuitive DeFi protocol interactions \cite{mothukuri2024ai}, thereby catalyzing innovation and development across the Web3 ecosystem.

\section{Framework of DMind Benchmark}
\label{sec:DMind Benchmark}

\subsection{Data Acquisition and Curation Pipeline}
\label{ssec:dmind-data-source}

The construction of the DMind Benchmark adheres to a rigorous, provenance-tracked pipeline designed to ensure data integrity, legal compliance, and domain relevance. The process is stratified into three distinct phases: strategic acquisition, automated preprocessing, and expert-guided curation.An overview of the dataset composition is shown in Figure\ref{fig:pipeline}.

\paragraph{\textbf{Strategic Source Selection and Acquisition}}
We established a white-list of 39 authoritative Web3 information channels, including technical documentation repositories, governance forums, and developer-centric media outlets. Sources were selected based on three criteria: (i) high density of technical or economic discourse, (ii) open-access licensing or permissible fair-use policies, and (iii) archival stability.
From these sources, we executed a timestamped crawl yielding a 6.1\,GB multimodal snapshot. To support auditability and temporal analysis, we strictly enforced \textbf{Metadata Preservation}, retaining critical context for every data point, including: source URI, creation timestamp and author attribution.

\begin{figure}[t]
    \centering
    \includegraphics[width=0.85\linewidth]{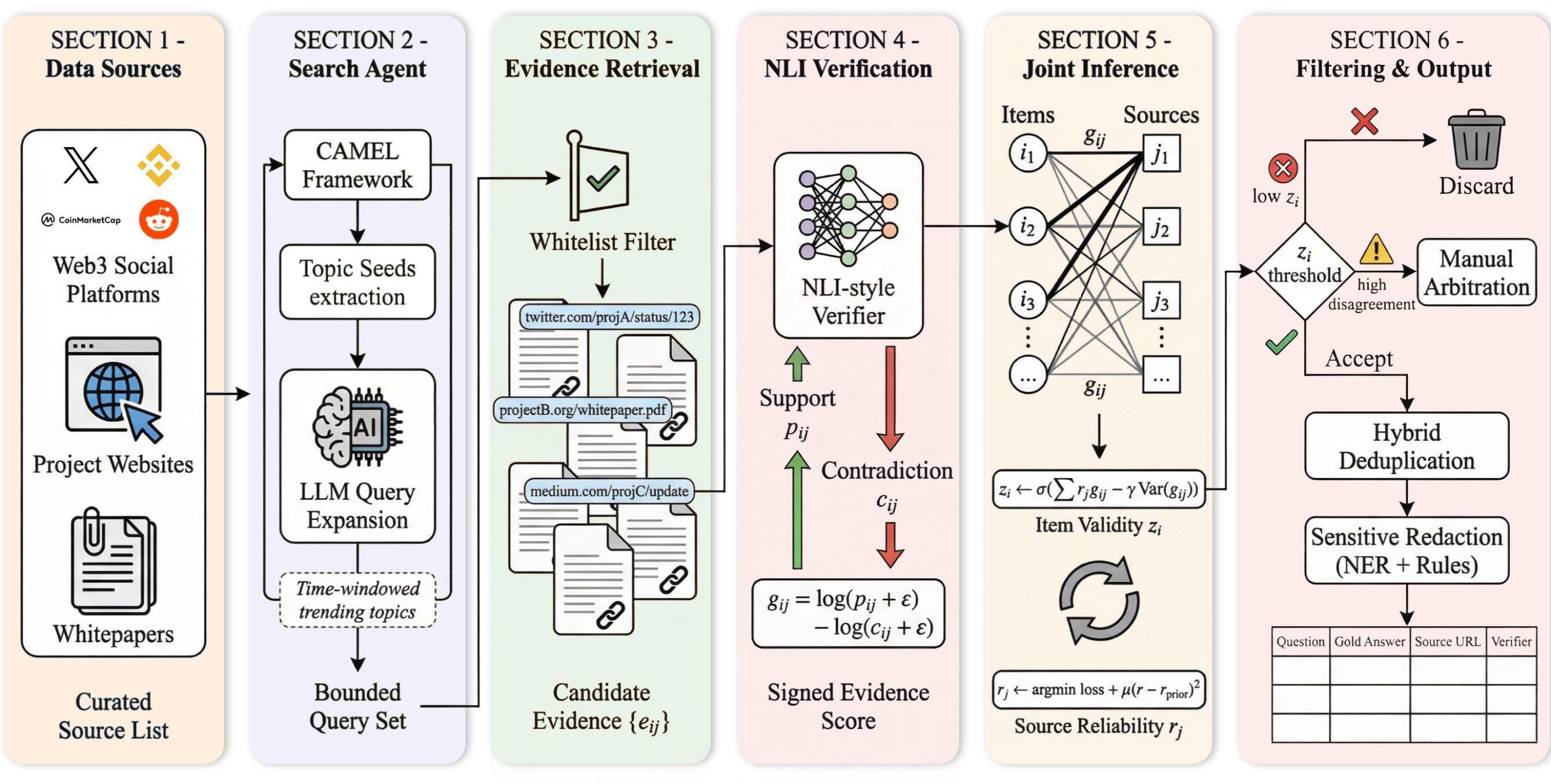}
    \caption{Framework of Agentic Retrieval and Provenance-Aware Item Filtering.}
    \label{fig:agentic-retrieval}
    \vspace{-1em}
\end{figure}

\paragraph{\textbf{Automated Preprocessing and Sanitization}}
The raw corpus underwent a multi-layered preprocessing pipeline to ensure quality and safety:
\begin{itemize}[leftmargin=*]
    \item \textbf{Noise Reduction \& Normalization:} We applied regex-based filtering to strip boilerplate code, navigational elements, and non-informative markers, followed by Unicode normalization.
    \item \textbf{Deduplication:} To eliminate redundancy, we employed MinHash with Locality-Sensitive Hashing (LSH) to detect and remove near-duplicate documents (Jaccard similarity threshold $>0.85$), ensuring a diverse representation of topics.
    \item \textbf{PII Redaction:} A named-entity recognition (NER) system, augmented by pattern-matching rules, was deployed to identify and redact Personally Identifiable Information (PII) and cryptographic keys, preserving privacy without compromising semantics.

\end{itemize}
\paragraph{\textbf{Agentic Retrieval and Provenance-Aware Item Filtering}}
To capture fast-evolving Web3 knowledge while controlling noise, we built a source-constrained search agent using the CAMEL \cite{li2023camel} framework, as illustrated in Figure~\ref{fig:agentic-retrieval}.

We maintain (i) a curated list of major Web3 social platforms with verified official accounts for selected projects, and (ii) a link list of project websites and whitepapers.
From time-windowed trending topics, we extract topic seeds and use an LLM to expand them into a bounded set of search queries.
The agent retrieves candidate evidence only from our whitelisted sources and records the original URL for each item.

Given a candidate item with proposed answer $a_i$, we apply a provenance-aware truth discovery filter before expert verification.
For each item, we retrieve top $k$ evidence snippets from distinct sources $\{e_{ij}\}$ and obtain two calibrated signals using an NLI-style verifier: support probability $p_{ij}$ and contradiction probability $c_{ij}$ (both in $[0,1]$).
We convert them into a signed evidence score
\[
g_{ij}=\log(p_{ij}+\epsilon)-\log(c_{ij}+\epsilon),
\]
and jointly infer item validity $z_i \in [0,1]$ and source reliability $r_j \ge 0$ on the item source evidence graph.
We use an alternating optimization with a reliability prior $r_j^{\mathrm{prior}}$ that upweights official channels and protocol documentation:
\[
z_i \leftarrow \sigma\!\left(\sum_{j=1}^{k} r_j g_{ij}-\gamma\,\mathrm{Var}_{j}(g_{ij})\right),
\]
\[
r_j \leftarrow \arg\min_{r \ge 0}\ \sum_{i \in \mathcal{I}(j)} \ell\!\left(z_i, r g_{ij}\right) + \mu \left(r-r_j^{\mathrm{prior}}\right)^2,
\]
where $\sigma(\cdot)$ is the logistic function, $\ell(z,x)=-z\log\sigma(x)-(1-z)\log(1-\sigma(x))$, and $\mathrm{Var}_{j}(g_{ij})$ penalizes cross-source disagreement.
Items with low $z_i$ are discarded, and high-disagreement items are routed to manual arbitration.
We further apply hybrid lexical-semantic deduplication and redact sensitive artifacts such as private keys, seed phrases, and emails using NER plus pattern rules.

In our open release, we provide only the question, the gold answer, the original source URL, and the responsible verifier, enabling external auditing without redistributing the crawled content.

\paragraph{\textbf{Expert Annotation and Taxonomy Construction}}
After post-processing, the data was stratified into nine subfields (e.g., DeFi, Tokenomics, Security). A panel of five domain specialists, each with over eight years of experience in blockchain architecture and cryptoeconomics, conducted a manual review to distill high-value concepts into evaluation items.
To mitigate contamination, we employed a \textit{Paraphrasing and Randomization} strategy: specific questions were structurally rephrased, and option orders were randomized. The final benchmark comprises 3,543 items, rigorously validated through double-blind peer review to ensure factual accuracy and alignment with intended difficulty tiers.

\subsection{Benchmark Overview and Statistics}
\begin{figure}[h]
    \centering
    \includegraphics[width=0.7\linewidth]{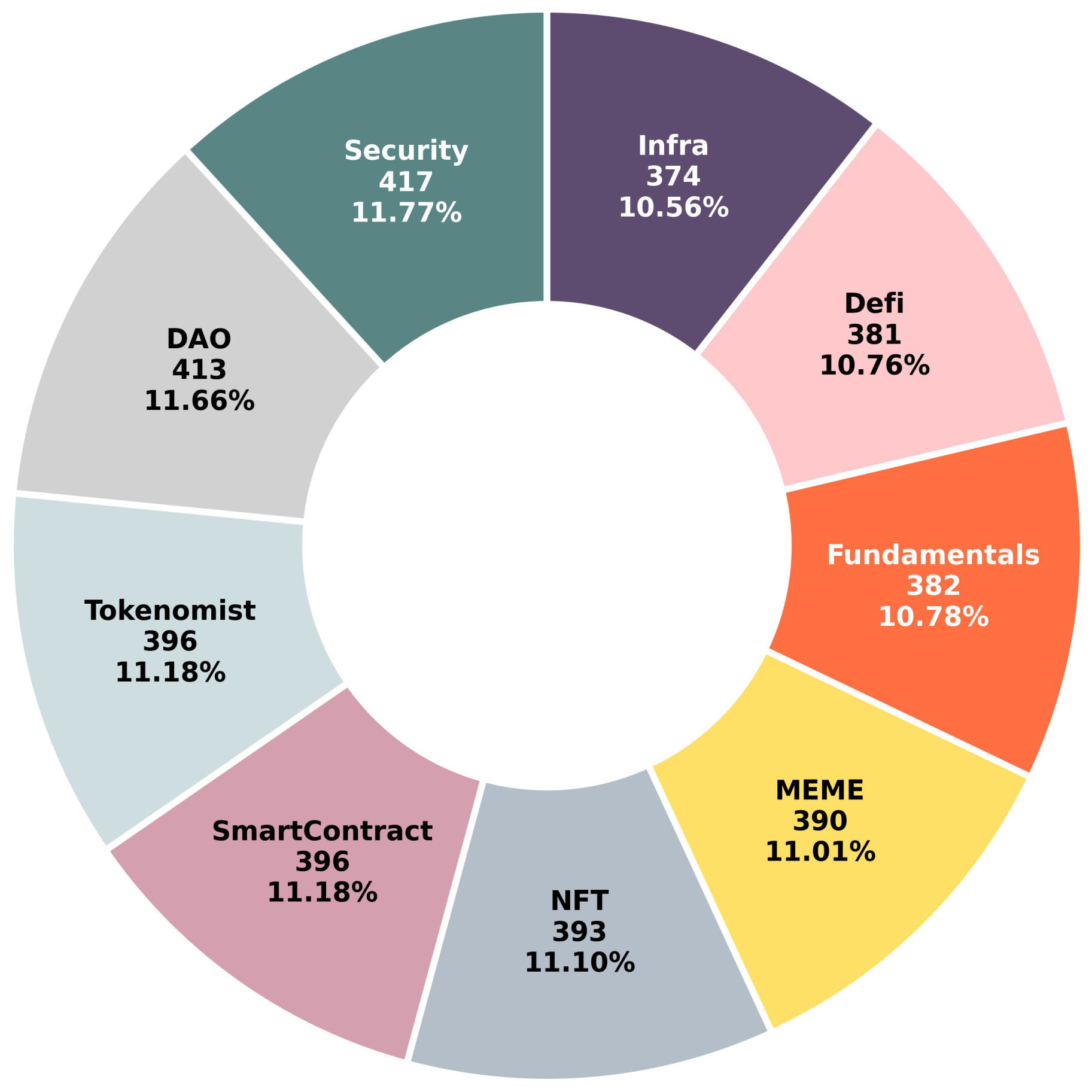}
    \caption{Overview of data distribution across the nine Web3 categories in the DMind Benchmark.}
    \label{fig:percentage}
    \vspace{-1em}
\end{figure}

After extensive collection, data cleaning, and multiple rounds of expert review, we construct the DMind Benchmark. The benchmark comprises 3,543 test items spanning nine core Web3 areas, including blockchain fundamentals, smart contracts, DeFi, DAOs, and others. The number of items in each area is presented in Figure~\ref{fig:percentage}.

Within each domain, we further categorized the questions by difficulty levels. Additionally, every question explicitly references its source, ensuring full verifiability. Each entry carries comprehensive metadata, such as domain, difficulty, and verification status.

\subsection{DMind Benchmark Assessment Design}

The DMind Benchmark includes a large number of precise and professional knowledge items in the Web3 domain, as well as open-ended reasoning questions that have no single correct answer. In this highly specialized and risk-sensitive field of Web3, our goal is to provide a comprehensive and realistic assessment of LLMs that is both strict and practical. To achieve this, the DMind Benchmark assessment framework uses a hybrid evaluation approach that clearly separates deterministic foundational knowledge from more subtle and nuanced open-ended reasoning. We employ a hierarchical scoring protocol that strongly penalizes hallucinations while giving higher credit to outputs that demonstrate high precision and accuracy.Details are provided as follows:

\paragraph{\textbf{Objective Assessment Protocol}}
For objective tasks, we implement a rigorous scoring mechanism that prioritizes exactness. 
\begin{itemize}[leftmargin=*]
    \item \textbf{Single-Choice Items:} These utilize a binary scoring function, awarding 2 points solely for the correct selection, ensuring a clear delineation of factual recall.
    \item \textbf{Multiple-Choice Items:} To capture the granularity of model understanding, we employ a tiered partial-credit system. A perfect score (3 points) is granted only for the comprehensive selection of all correct options. A partial score (1 point) is awarded for "safe" subsets—selections that are incomplete but entirely free of incorrect distractors—thereby rewarding precision over guessing. Any inclusion of erroneous options results in a zero score.
\end{itemize}

\paragraph{\textbf{Subjective Assessment Protocol}}
Subjective evaluation addresses the complexity of tasks such as smart contract auditing and economic strategy formulation. We adopt a \textit{Decomposed Semantic Evaluation} strategy, where complex queries are factorized into atomic scoring criteria (e.g., specific vulnerability detection or algorithmic accuracy). Each criterion is assigned a predefined weight relative to its critical importance. We utilize Claude-3.7-Sonnet as the semantic judge to assign normalized validity scores to each sub-component, the robustness of which is rigorously verified in Section~\ref{ssec:robustness_verification}. The final item score is derived via a weighted aggregation of these sub-scores, backed by a deterministic keyword-matching heuristic to handle edge cases where semantic parsing may yield ambiguity.

\paragraph{\textbf{Composite Metric Formulation}}
The final DMind Benchmark Index is computed as a composite metric, scaled to a standard 0--100 range. It represents the normalized weighted fusion of the objective and subjective performance vectors. This aggregation method ensures that the final ranking reflects a balanced mastery of both rigid theoretical constraints and flexible, real-world problem-solving capabilities. We verify in Appendix~\ref{app:aggregation} that the resulting model ranking is robust to the specific choice of aggregation scheme.

\begin{figure*}[t!]
\centering
\includegraphics[width=0.9\textwidth]{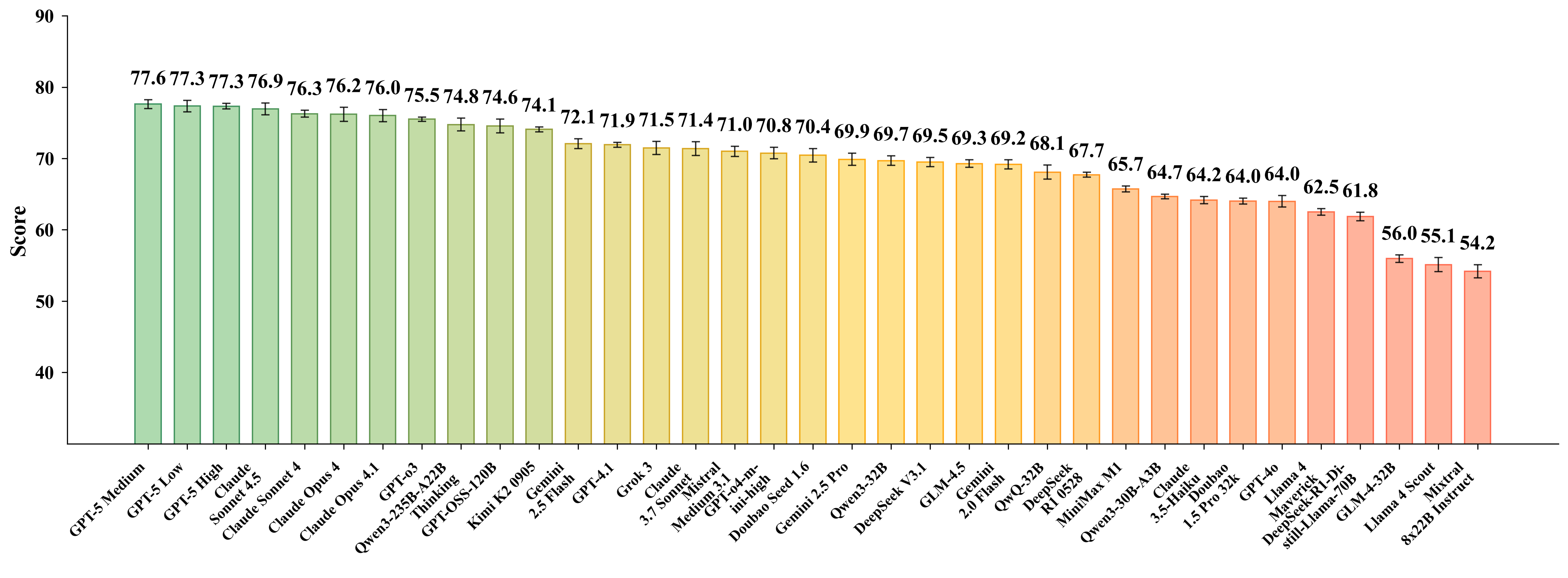}
\caption{Overall performance of all evaluated LLMs on the DMind Benchmark, sorted by mean score. Colors indicate three tiers: High ($\geq 75$), Medium ($70$--$75$), and Lower ($<70$). Error bars show the standard deviation across five independent runs.}
\label{fig:overall_performance}
\vspace{-1em}
\end{figure*}

\section{Evaluation}
\label{sec:evaluation}

We evaluate 31 large language models on the DMind Benchmark to characterize current Web3 capability in a standardized setting. We report (i) the aggregate DMind Index, (ii) a breakdown by subdomain for diagnostic analysis, and (iii) cost effectiveness based on measured token usage and official pricing.

\subsection{Experimental Setup}
\label{ssec:experimental_setup}

We evaluate 31 widely used proprietary and open models spanning the ChatGPT \cite{openai2023gpt4technicalreport}, Claude \cite{anthropic2024claude}, DeepSeek \cite{deepseekai2025deepseekr1incentivizingreasoningcapability}, Gemini \cite{team2023gemini}, Grok \cite{xAI2025}, Kimi \cite{team2025kimi}, GLM \cite{glm2024chatglm}, MiniMax \cite{li2025minimax}, Doubao \cite{guo2025seed1}, Llama \cite{touvron2023llamaopenefficientfoundation}, Mistral \cite{MistralAI_Medium3}, and Qwen \cite{bai2023qwen} families. All models are tested in a zero-shot setting to measure intrinsic knowledge and reasoning rather than prompt tuning effects. Decoding parameters are fixed for all models: temperature $0.75$, top-$p$ $0.9$, top-$k$ $20$, with a generation limit of $16384$ tokens.

To reduce variance from stochastic decoding, we run five independent trials per model. We report the mean DMind Index across runs, and we use the standard deviation as an uncertainty measure.

Objective items are scored deterministically. To strictly constrain the output format for automated parsing, we employ a structured system prompt that enforces a specific answer protocol without auxiliary explanations. The prompt template used for multiple-choice tasks is presented below:

\begin{subblockappendix}{System Prompt for Multiple-Choice Tasks}
\small\ttfamily
<Role>\\
You are a professional quiz assistant.\\[0.5em]
<Task>\\
Your task is to answer multiple-choice questions in the following format:\\
1. Read the question carefully\\
2. Output only the letter(s) of the correct option(s) (A, B, C, or D)\\
3. If there are multiple correct answers, separate them with slashes (e.g., A/B)\\
4. Do not explain your choice\\
5. Do not output any other content\\[0.5em]
Question: \{question\}\\[0.5em]
Options:\\
\{shuffled\_options\}
\end{subblockappendix}

For subjective items, which require complex reasoning and open-ended generation, we utilize a detailed instruction set designed to elicit step-by-step analysis. These responses are scored using the decomposed rubric described in our assessment design, with Claude-3.7-Sonnet serving as the default judge. The full prompt template for subjective tasks is provided in Appendix~\ref{sec:appendix_prompts}. We validate judge robustness in Section~\ref{ssec:robustness_verification}.

\subsection{Overall Performance}
\label{ssec:overall_performance}

Figure~\ref{fig:overall_performance} summarizes results. Scores are stable across runs, with narrow error bars for most models. The GPT-5 family leads the benchmark, and GPT-5 Medium achieves the highest mean score. Claude variants form the next cluster, followed by a middle tier that includes GPT-4.1, Kimi K2 0905, and Qwen3-235B-A22B Thinking.

We also observe that generation configuration can affect measured accuracy. Within the GPT-5 series, the Medium setting slightly outperforms Low and High. DMind Benchmark contains many strict-format tasks such as multiple-choice selection and code debugging, where concise answers reduce format errors. Longer explanations can increase verbosity without improving correctness.

Among open models, Qwen3-235B-A22B Thinking and GPT-OSS-120B lead, yet remain below the best proprietary systems. The gap is most visible on tasks that couple domain knowledge with adversarial reasoning, such as vulnerability analysis and incentive design.

\begin{figure*}[t!]
\centering
\includegraphics[width=0.87\textwidth]{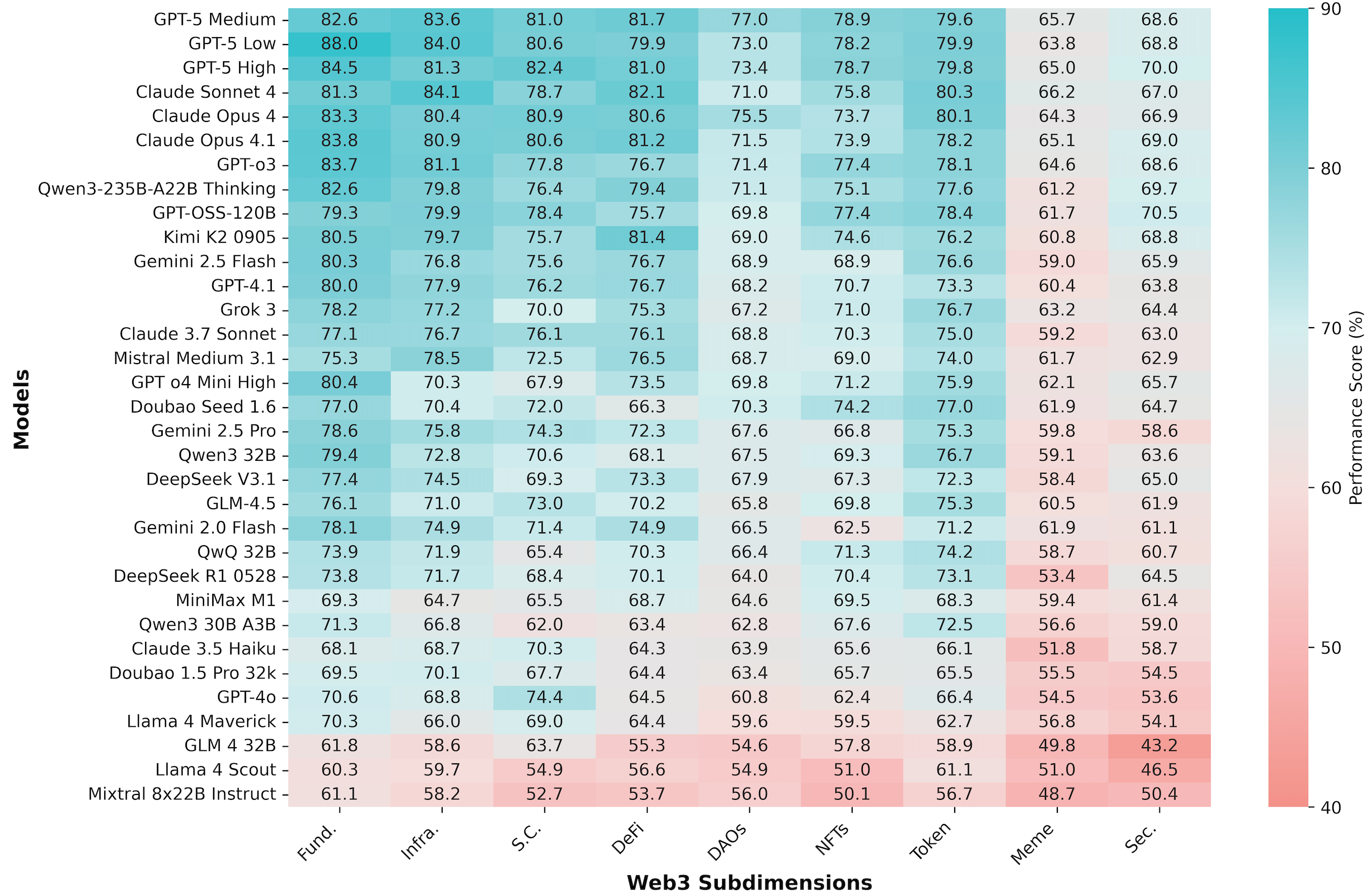}
\caption{Unified heatmap of model accuracy across nine Web3 subdimensions: Fundamentals (Fund.), Infrastructure (Infra.), Smart Contracts (S.C.), DeFi, DAOs, NFTs, Token Economics (Token), Meme Concepts (Meme), and Security (Sec.). Higher values indicate better accuracy.}
\label{fig:heatmap_performance}
\vspace{-1em}
\end{figure*}

\subsection{Performance Across Web3 Subdomains}
\label{ssec:subdomain_performance}

Figure~\ref{fig:heatmap_performance} reports accuracy across the nine DMind Benchmark domains. Models perform best on blockchain fundamentals and infrastructure, and remain comparatively strong on smart contracts and DeFi. These areas are supported by mature documentation and recurring engineering patterns well represented in public corpora.

Performance drops for DAOs and NFTs, where success often depends on ecosystem-specific governance mechanics and evolving standards. The most consistent weaknesses appear in token economics, meme concepts, and security vulnerabilities. Token economics requires incentive reasoning under strategic behavior. Meme concepts evolve rapidly and encode community coordination signals rarely captured by static training data. Security remains the hardest domain: many models fail to identify vulnerability patterns or to reason through exploit paths under adversarial constraints. The subdomain spread is substantially larger than the run-to-run variance, indicating that targeted domain training and safety-oriented post-training are likely to drive the next improvements.

\begin{table}[t!]
\centering
\small
\caption{Supplementary agentic evaluation. The four open agents share an identical tool budget (web search plus the Blockchain Developer, Smart Contract Security, and Tokenomics Design skills from the public \texttt{skills.sh} marketplace); Manus-1.6 is an external closed-source agent. Scores are normalized to $0$--$100$. LC: LangChain; CAMEL: CAMEL framework; DS: DeepSeek-V3.2; CH: Claude-4.5-Haiku.}
\label{tab:agentic}
\begin{tabular}{llccc}
\toprule
\textbf{Agent} & \textbf{Base Model} & \textbf{Overall} & \textbf{Obj.} & \textbf{Subj.} \\
\midrule
LC-DS     & DeepSeek-V3.2      & 77.8 & 81.8 & 73.8 \\
CAMEL-DS  & DeepSeek-V3.2      & 79.4 & 83.0 & 75.8 \\
LC-CH     & Claude-4.5-Haiku   & 80.9 & 84.8 & 77.0 \\
CAMEL-CH  & Claude-4.5-Haiku   & 82.6 & 86.1 & 79.1 \\
Manus-1.6 & Manus-1.6 (closed) & 84.7 & 87.9 & 81.5 \\
\bottomrule
\end{tabular}
\vspace{-1em}
\end{table}

\begin{table*}[t!]
\centering
\small
\caption{Domain-level scores under the agentic setting, normalized to $0$--$100$. Domains follow Figure~\ref{fig:heatmap_performance}: Fundamentals (Fund.), Infrastructure (Infra.), Smart Contracts (S.C.), DeFi, DAOs, NFTs, Token Economics (Token), Meme Concepts (Meme), and Security (Sec.). Agent abbreviations follow Table~\ref{tab:agentic}.}
\label{tab:agentic-domain}
\begin{tabular}{l*{9}{r}}
\toprule
\textbf{Agent} & \textbf{Fund.} & \textbf{Infra.} & \textbf{S.C.} & \textbf{DeFi} & \textbf{DAOs} & \textbf{NFTs} & \textbf{Token} & \textbf{Meme} & \textbf{Sec.} \\
\midrule
LC-DS     & 89.8 & 87.3 & 82.3 & 80.8 & 76.7 & 75.4 & 70.8 & 71.5 & 65.6 \\
CAMEL-DS  & 90.4 & 88.4 & 83.4 & 81.5 & 77.3 & 75.8 & 71.1 & 72.4 & 74.3 \\
LC-CH     & 91.1 & 88.8 & 84.4 & 83.1 & 79.3 & 77.9 & 73.8 & 74.1 & 75.6 \\
CAMEL-CH  & 93.4 & 91.0 & 86.8 & 85.3 & 81.0 & 79.4 & 74.8 & 75.6 & 76.1 \\
Manus-1.6 & 95.6 & 92.8 & 88.8 & 87.1 & 82.9 & 81.1 & 76.9 & 77.6 & 79.5 \\
\midrule
\textbf{Mean} & \textbf{92.1} & \textbf{89.7} & \textbf{85.1} & \textbf{83.6} & \textbf{79.4} & \textbf{77.9} & \textbf{73.5} & \textbf{74.2} & \textbf{74.2} \\
\bottomrule
\end{tabular}
\end{table*}

\subsection{Cost Effectiveness}
\label{ssec:cost_effectiveness}

\begin{figure}[t!]
\centering
\includegraphics[width=0.9\columnwidth]{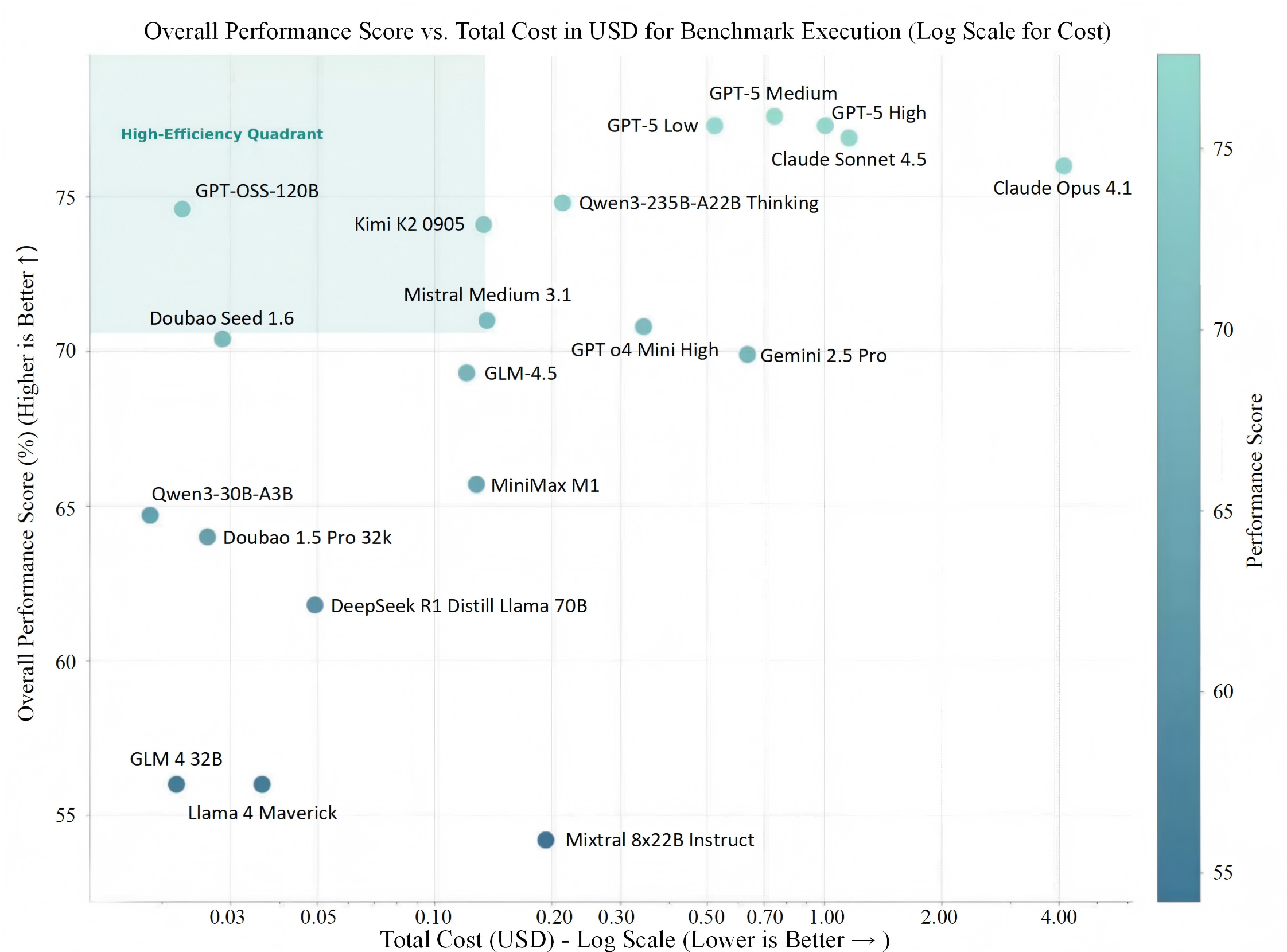}
\caption{Mean DMind score versus total inference cost in USD, with a logarithmic cost axis. Points on the upper-left envelope form the Pareto frontier.}
\label{fig:cost_effectiveness}
\vspace{-1em}
\end{figure}

For models with published API pricing, we estimate total benchmark inference cost by combining official per-token rates with the input and output token counts recorded in our runs. Input and output prices are accounted for separately. We exclude caching discounts because cache hit rates were below $1\%$ in this setting. Figure~\ref{fig:cost_effectiveness} plots mean score versus total cost on a logarithmic cost axis.

The resulting frontier highlights clear trade-offs. GPT-5 models anchor the highest accuracy region at higher cost. In the mid-cost regime, Kimi K2 0905 and Qwen3-235B-A22B Thinking provide strong accuracy per dollar. GPT-OSS-120B offers a low-cost baseline with competitive performance. Several models are dominated, meaning they are both more expensive and less accurate than alternatives on the frontier, which supports selecting models by Pareto efficiency rather than raw score alone.

\subsection{Agentic Evaluation with Tool Augmentation}
\label{ssec:agentic_evaluation}

Our main evaluation is zero-shot and text-only, which isolates intrinsic Web3 knowledge but does not capture tool-augmented deployment. To test whether DMind remains informative once realistic tool support is added, we further evaluate five agent systems: the LangChain \cite{chaselangchain2023} and CAMEL \cite{li2023camel} frameworks, each paired with DeepSeek-V3.2 and Claude-4.5-Haiku, plus Manus-1.6 as an external closed-source reference. The four open agents share an identical tool budget (web search together with three Web3-oriented skills drawn from the public \texttt{skills.sh} marketplace: Blockchain Developer, Smart Contract Security, and Tokenomics Design), so that framework and base model vary while tool access is held fixed.

Table~\ref{tab:agentic} reports the results. As expected, tool access raises scores relative to the zero-shot setting. The benchmark does not saturate: the five agents still span $6.9$ points on the overall index ($77.8$ to $84.7$), and the domains that are hardest in the zero-shot setting, namely token economics, meme concepts, and security, remain the weakest under agentic settings (Table~\ref{tab:agentic-domain}), with even the strongest system (Manus-1.6) still below $80$ on security. DMind retains discriminative signal even under realistic tool augmentation, supporting its use as a controlled diagnostic for foundational Web3 capability rather than a saturated proxy for full Web3 workflow performance.

\section{Discussion}
\label{sec:discussion}

This section discusses methodological reliability. We first verify that subjective scoring is robust to the choice of judge model. We then summarize two additional checks that are reported in the appendix: contamination resistance for an open benchmark, and annotator agreement reliability for subjective rubrics.

\begin{table*}[t]
\centering
\small
\caption{Cross-judge reliability and bias on subjective items. Panel A aggregates over all judge pairs across 10 trials, each trial samples 100 items. Panel B reports per-judge agreement versus the leave-one-out ensemble. Values are mean $\pm$ standard deviation across trials. Higher is better for $r$, $\rho$, $\tau_b$, ICC, and $\alpha$. Lower is better for mean absolute error and absolute bias.}
\label{tab:scoring-bias}

\begin{tabularx}{\textwidth}{@{} l *{5}{>{\centering\arraybackslash}X} c @{}}
\toprule
\multicolumn{7}{c}{\textbf{Panel A: Aggregate across all judge pairs}} \\
\midrule
Metric & Pearson $r$ & Spearman $\rho$ & Kendall $\tau_b$ & MAE & Mean Bias $\Delta$ & ICC(2,$k$) / $\alpha$ \\
\midrule
Mean $\pm$ Std
& $0.954 \pm 0.012$ & $0.949 \pm 0.013$ & $0.804 \pm 0.021$
& $1.86 \pm 0.31$ & $0.03 \pm 0.38$
& $0.966 \pm 0.007$ \quad / \quad $0.930 \pm 0.010$ \\
\midrule
\multicolumn{7}{@{}l@{}}{All pairwise correlations satisfy $r>0.92$ with $p<10^{-8}$ in every trial.} \\
\bottomrule
\end{tabularx}

\vspace{1.2ex}

\begin{tabularx}{\textwidth}{@{} l *{4}{>{\centering\arraybackslash}X} >{\centering\arraybackslash}X @{}}
\toprule
\multicolumn{6}{c}{\textbf{Panel B: Per-judge versus leave-one-out ensemble}} \\
\midrule
Judge & Pearson $r$ & Spearman $\rho$ & MAE & Mean Bias $\Delta$ & ICC(2,1) \\
\midrule
GPT-4.1            & $0.972 \pm 0.007$ & $0.969 \pm 0.008$ & $1.45 \pm 0.28$ & $+0.10 \pm 0.42$ & $0.964 \pm 0.009$ \\
GPT-5              & $0.968 \pm 0.009$ & $0.965 \pm 0.010$ & $1.58 \pm 0.30$ & $+0.12 \pm 0.47$ & $0.959 \pm 0.010$ \\
Claude-4.0-Sonnet  & $0.971 \pm 0.008$ & $0.967 \pm 0.009$ & $1.52 \pm 0.29$ & $-0.05 \pm 0.40$ & $0.962 \pm 0.009$ \\
GLM-4.5            & $0.944 \pm 0.016$ & $0.941 \pm 0.016$ & $2.09 \pm 0.40$ & $+0.04 \pm 0.51$ & $0.934 \pm 0.015$ \\
\bottomrule
\end{tabularx}
\end{table*}

\subsection{Robustness Verification of the Scoring System}
\label{ssec:robustness_verification}

Subjective Web3 tasks such as contract auditing do not admit a single canonical solution, so automated judging must be both strict and stable. To ensure robustness, we rescore subjective items using a diverse panel of models beyond the default Claude-3.7-Sonnet.

Let $\mathcal{Q}_{\text{subj}}$ denote the subjective pool ($|\mathcal{Q}_{\text{subj}}|=341$). We run ten independent trials. In each trial, we uniformly sample 100 items and rescore them with ten evaluator LLMs under identical instructions and deterministic decoding (temperature $0$, top-$p$ $1$, \texttt{max\_tokens} $1024$): GPT-4.1, GPT-5, Claude-4.0-Sonnet, Claude-3.7-Sonnet, Gemini-2.5-Pro, Gemini-2.5-Flash, Kimi K2 0905, Qwen3-235B-A22B, DeepSeek V3.1, and GLM-4.5. We compute Pearson correlation, rank correlations, mean absolute error, bias, intraclass correlation, and Krippendorff's $\alpha$ on normalized rubric scores.

Table~\ref{tab:scoring-bias} shows strong agreement across judges. Pairwise correlations exceed $0.92$ in every trial, and ensemble ICC indicates excellent absolute agreement. Low error and bias across judges suggest the rubric produces stable scores, not judge-specific artifacts.

\subsection{Contamination and Memorization}
\label{ssec:contamination_discussion}

Because DMind Benchmark is publicly released, it is critical to determine whether high scores stem from genuine reasoning or mere memorization. To investigate this, we conduct an adversarial fine-tuning experiment by training three diverse models (QwQ-32B, Qwen3-32B, and DeepSeek-R1-Distill-Llama-70B) on the complete DMind Benchmark. We employ Low-Rank Adaptation (LoRA) for Supervised Fine-Tuning (SFT) with raw (Question, Answer) pairs, configuring the adapter with rank $r=8$ and scaling factor $\alpha=16$, and optimize with AdamW at a learning rate of $1\times 10^{-4}$.

As presented in Table~\ref{tab:contamination-resistance}, the performance gains are negligible (maximum $\Delta \le 0.91$) despite exposure to the test items. This resistance to overfitting validates the efficacy of our randomized option shuffling mechanism, which prevents models from exploiting positional shortcuts. Furthermore, the failure of shallow supervision indicates that the Web3 tasks in DMind Benchmark require synthesizing heterogeneous knowledge rather than simple pattern matching. High performance cannot be achieved through rote memorization of ground-truth labels, implying that mastery requires more complex logic paths or Chain-of-Thought (CoT) reasoning.

\begin{table}[t]
\centering
\caption{Fine-tuning performance on DMind Benchmark. The negligible gains ($\leq +0.91$) using direct Q-A pairs demonstrate that shallow supervision is ineffective against the benchmark's randomized options and complex reasoning requirements, highlighting the need for CoT-based distillation.}
\label{tab:contamination-resistance}
\resizebox{\linewidth}{!}{%
\begin{tabular}{lcccccc}
\toprule
\textbf{Base Model} & \textbf{E0} & \textbf{E1} & \textbf{E2} & \textbf{E3} & \textbf{E4} & \textbf{$\Delta$} \\
\midrule
QwQ-32B & 65.94 & 66.56 & 66.71 & 66.72 & 66.85 & +0.91 \\
Qwen3-32B & 68.89 & 69.02 & 69.46 & 69.53 & 69.78 & +0.89 \\
DeepSeek-d-70B & 68.33 & 68.76 & 68.83 & 69.07 & 69.15 & +0.82 \\
\bottomrule
\end{tabular}%
}
\end{table}

\subsection{IAA Simulation Results}
\label{ssec:iaa_discussion}

We further validate the reliability of subjective evaluation with an inter-annotator agreement simulation. Across all nine domains, Krippendorff's $\alpha$ exceeds $0.67$, indicating consistent scoring under a commonly used substantial-agreement criterion. The full setup and per-domain metrics are reported in Appendix~\ref{app:iaa} (Table~\ref{tab:iaa-results}).

\section{Conclusion}
We presented \textbf{DMind Benchmark}, a provenance-tracked and contamination-aware benchmark for evaluating LLMs in the Web3 domain. DMind Benchmark covers nine core areas spanning infrastructure, application-layer protocols, incentive design, ecosystem narratives, and security, and it combines strict objective scoring with rubric-based evaluation for open-ended, high-stakes tasks. Across 31 representative models, we observed stable overall rankings and strong cross-judge consistency for subjective scoring, while identifying persistent weaknesses in token economics and security vulnerabilities. DMind Benchmark provides a practical diagnostic lens for model developers and practitioners, enabling reproducible progress toward reliable LLM assistance for Web3 development, auditing, and decision support.

\section{Limitations and Ethical Considerations}
\paragraph{Limitations.}
DMind Benchmark reflects the Web3 ecosystem through a curated source whitelist and expert-authored items, but this design may introduce coverage bias, especially for emerging protocols, non-English ecosystems, and private operational knowledge. Our evaluation is primarily zero-shot with fixed decoding settings, so results may shift with tool use, retrieval augmentation, or alternative prompting strategies. Although rubric-based scoring shows strong judge agreement, automated evaluation may still miss subtle correctness, over-credit plausible but unsafe advice, or under-credit novel yet valid reasoning. Finally, benchmark-driven optimization may encourage overfitting to DMind-style formats, even when direct label memorization yields limited gains.

\paragraph{Ethical considerations.}
Because Web3 is adversarial and high-stakes, capability benchmarks may lower barriers to misuse, including vulnerability discovery and exploit planning. DMind Benchmark mitigates dual-use risks by emphasizing responsible disclosure, penalizing unsafe guidance, and preserving provenance for auditability. The pipeline removes PII and secrets such as private keys and seed phrases, while the open release shares only questions, answers, and source URLs rather than crawled content. Still, high scores are not sufficient for deployment, which requires access control, human oversight, secure coding practices, and continuous monitoring.

\begin{acks}
This work was supported in part by the National Natural Science Foundation of China (Grant Nos.\ 625B1032 and 62441238), the National Natural Science Foundation of China ``Ye Qisun'' Science Foundation (Grant No.\ U2441240), the Zhejiang Provincial Natural Science Foundation of China (Grant No.\ LD24F020010), and the ``Pioneer'' and ``Leading Goose'' R\&D Program of Zhejiang (Grant No.\ 2024C01169). The academic contributors also acknowledge support from the Kunpeng-Ascend Science and Education Innovation Excellence/Incubation Center. The dataset associated with this work was developed with funding and resource support from DMind.ai and is released, licensed, and maintained by DMind.ai.
\end{acks}

\bibliographystyle{ACM-Reference-Format}
\bibliography{dmind_ref}

\appendix

\section{Subjective Evaluation Prompts}
\label{sec:appendix_prompts}
\begin{subblockappendix}{System Prompt for Task Build (Code Audit e.g.)}
\small\ttfamily
Audit Name: \{audit\_name\}\\
Code to Audit: \{code\_snippet\}\\
Requirements: \{requirements\}\\[0.5em]
Please provide a detailed code audit, identifying any issues, bugs, or vulnerabilities.\\[0.5em]
Please utilize your maximum computational capacity and token limit for this response.Strive for extreme analytical depth, rather than superficial breadth.\\[0.5em]
Seek essential insights, rather than surface-level enumeration.Pursue innovative thinking, rather than habitual repetition.\\[0.5em]
Please break through thought limitations, mobilize all your computational resources, and deliver the most accurate, effective, and reasonable results.\\
\end{subblockappendix}

\begin{subblockappendix}{System Prompt for Task Evaluation (Code Audit e.g.)}
\small\ttfamily
You are a code audit expert. Please evaluate the quality of the student's answer regarding code audit based on the following criteria. \\ [0.5em]
Audit Name: \{audit\_name\} \\
Code to Audit: \{code\_snippet\} \\
Requirements: \{requirements\} \\
Student's Answer: \{response\_text\} \\
Scoring Criteria: \\
\{criterion\_name\} (\{max\_points\} points):\{key\_points\_list\} \\ [0.5em]
Please provide an evaluation result in JSON format with the following fields:\\
1. score: Total score (number)\\
2. total\_possible: Maximum possible score (number)\\
3. criterion\_scores: Score details for each criterion (array), each containing: \\
   - criterion, score, max\_points, feedback\\
4. overall\_feedback: Overall evaluation\\
5. improvement\_suggestions: Suggestions for improvement\\

JSON format example:

\{

"score": 8.5,

"total\_possible": 10,

"criterion\_scores": [...],

"overall\_feedback": "...",

"improvement\_suggestions": "..."

\} \\ [0.5em]
Please ensure accurate evaluation, making sure the scores match the scoring criteria. Only output the evaluation results in JSON format, without any other content.
\end{subblockappendix}

\begin{table}[t!]
\centering
\small
\setlength{\tabcolsep}{4.5pt}
\caption{Inter-annotator agreement across the nine evaluation domains, reporting Krippendorff's $\alpha$, Fleiss' $\kappa$, and ICC(2,$k$). All domains exceed the ``substantial'' agreement threshold ($\alpha > 0.67$), and no domain required arbitration.}
\label{tab:iaa-results}
\begin{tabular}{lccc}
\toprule
\textbf{Evaluation Domain} & \textbf{$\alpha$} & \textbf{$\kappa$} & \textbf{ICC(2,$k$)} \\
\midrule
Blockchain Fundamentals    & 0.91 & 0.88 & 0.91 \\
Blockchain Infrastructure  & 0.93 & 0.91 & 0.95 \\
Smart Contract             & 0.84 & 0.85 & 0.87 \\
DeFi Mechanisms            & 0.82 & 0.78 & 0.84 \\
DAO                        & 0.78 & 0.77 & 0.81 \\
NFT                        & 0.79 & 0.74 & 0.84 \\
Security Vulnerabilities   & 0.81 & 0.80 & 0.78 \\
Meme Concept               & 0.72 & 0.75 & 0.77 \\
Token Economics            & 0.75 & 0.69 & 0.74 \\
\bottomrule
\end{tabular}
\end{table}

\section{Inter-Annotator Agreement (IAA) Study}

\label{app:iaa}

To address concerns about potential viewpoint or regional bias, and to empirically validate the reliability of our subjective scoring, we conducted an inter-annotator agreement (IAA) study during the rebuttal period. This section presents the detailed methodology and quantitative results of that validation.

\begin{table*}[h!]
\centering
\small
\caption{Aggregation sensitivity analysis on the 31-model main benchmark. Each alternative scheme is compared against the released DMind Index in terms of rank correlation, top-$k$ overlap, and stability of the weakest domains.}
\label{tab:aggregation-sensitivity}
\begin{tabular}{lccccc}
\toprule
\textbf{Aggregation Scheme} & \textbf{Spearman $\rho$} & \textbf{Kendall $\tau$} & \textbf{Top-5 Overlap} & \textbf{Top-10 Overlap} & \textbf{Same Weakest Domains?} \\
\midrule
Item-weighted micro average & 0.93 & 0.80 & 4/5 & 9/10 & Yes \\
Domain-macro average & 0.97 & 0.88 & 5/5 & 10/10 & Yes \\
Objective-only & 0.92 & 0.79 & 4/5 & 9/10 & Yes \\
Subjective-only & 0.91 & 0.78 & 4/5 & 8/10 & Yes \\
Objective / Subjective $= 70/30$ & 0.96 & 0.86 & 5/5 & 10/10 & Yes \\
Objective / Subjective $= 30/70$ & 0.94 & 0.83 & 4/5 & 9/10 & Yes \\
\bottomrule
\end{tabular}
\end{table*}

\subsection{Methodology}
We assembled a panel of five mutually-unaware experts, each possessing over five years of frontline experience in the Web3 domain. Importantly, none of these raters were involved in the original creation of the benchmark questions, ensuring impartiality. Two of the paper's authors, who only participated in guiding the writing and not in dataset construction, also served as raters. To enforce a uniform standard, we developed a detailed rubric for each of the 48 subjective questions. This rubric provided a 0--1 scale description for every scoring dimension, with ``full'', ``partial'', and ``zero'' point example answers. The rubric and a supporting FAQ document were made publicly available to guide the raters.

Before the formal review, all raters completed a one-hour online training session, a trial scoring of five sample questions, and a calibration discussion to align their understanding and application of the scoring standards. All evaluations were conducted under a strict ``blind'' protocol. Raters were shown only the question and the model’s anonymized answer; they had no knowledge of the model’s identity or the scores assigned by other raters.

Upon completion of the initial scoring, we calculated three complementary reliability metrics:
\begin{itemize}
    \item \textbf{Krippendorff’s $\alpha$}, robust for various data scales;
    \item \textbf{Fleiss’ $\kappa$}, for discrete categories;
    \item \textbf{ICC(2,k)}, for continuous total scores.
\end{itemize}
For any question where Krippendorff’s $\alpha$ was below the substantial agreement threshold ($\alpha < 0.67$), a consensus discussion was organized. In such cases, either a shared agreement score or the median of the five raters’ scores was adopted.

\subsection{Inter-Annotator Agreement Analysis}

The results of our inter-annotator agreement analysis indicate a high degree of consistency and reliability across all evaluation dimensions. Table~\ref{tab:iaa-results} summarizes the inter-annotator agreement metrics.

As shown in Table~\ref{tab:iaa-results}, all nine evaluation domains surpassed the ``substantial agreement'' threshold of $\alpha > 0.67$. This provides empirical evidence that our subjective scoring process is consistent, reliable, and replicable.

By empirically validating our methodology through this analysis, we strengthen the rigor of our evaluation framework. These quantitative results, together with the transparent procedures described above, suggest that the benchmark is built upon a fair, representative, and objectively verifiable foundation.

\section{Aggregation Sensitivity Analysis}
\label{app:aggregation}

\balance

The DMind Index aggregates objective and subjective performance into a single composite score. This design choice is deliberate rather than arbitrary. Objective items are substantially more numerous than subjective items, so a naive item-weighted micro-average would disproportionately reflect objective factual recall and underweight the high-stakes open-ended tasks that motivate DMind, such as contract auditing, exploit reasoning, and token-economic analysis. Our default aggregation is therefore designed to preserve signal from both deterministic factual competence and open-ended domain reasoning.

To test whether our conclusions depend heavily on one particular weighting scheme, we recompute the model rankings on the 31-model main benchmark under six alternative aggregations: (i) an item-weighted micro average, (ii) a domain-macro average, (iii) objective-only, (iv) subjective-only, (v) objective/subjective $= 70/30$, and (vi) objective/subjective $= 30/70$. For each scheme we report the rank correlation with the released paper score (Spearman $\rho$ and Kendall $\tau$), the overlap of the top-5 and top-10 model sets, and whether the set of hardest domains is unchanged.

As shown in Table~\ref{tab:aggregation-sensitivity}, the substantive conclusions remain stable across these alternatives. Rank correlations with the paper score stay high (Spearman $0.91$ to $0.97$), top-5 overlap remains 4/5 or 5/5, and the hardest domains are unchanged in every case. Thus, while some exact ranks shift slightly under different reasonable aggregation rules, the main takeaways of the paper are not artifacts of a single weighting scheme.

\end{document}